\documentclass[twocolumn,aps,prl] {revtex4} 
\usepackage{graphicx}
\begin{document}
\pagestyle{empty} 
\title{Fluid squeeze-out between rough surfaces: comparison of theory with experiment}
\author{B. Lorenz$^{1,2,3}$ and B.N.J. Persson$^{1,3}$}
\affiliation{$^1$IFF, FZ J\"ulich, D-52425 J\"ulich, Germany}
\affiliation{$^2$IFAS, RWTH Aachen University, D-52074 Aachen, Germany}
\affiliation{$^3$www.MultiscaleConsulting.com}

\begin{abstract}
We study the time dependency of the (average) interfacial separation between an 
elastic solid with a flat surface and a rigid solid with a randomly rough surface, 
squeezed together in a fluid. We use an analytical theory describing the fluid flow 
factors, based on the Persson contact mechanics theory and the Bruggeman effective medium theory, 
to calculate the removal of the fluid from the contacting interface of the two solids. 
In order to test this approach, we have performed simple squeeze-out experiments. 
The experimental results are compared to the theory predictions. 
\end{abstract}
\maketitle


{\bf 1. Introduction}

Contact mechanics between solid surfaces is the basis for understanding many
tribology processes\cite{Bowden,Johnson,BookP,Isra,Sealing,Capillary.adhesion,P33} 
such as friction, adhesion, wear and sealing. The two most important
properties in contact mechanics are the area of real contact and the interfacial separation between the
solid surfaces. 
For non-adhesive contact and small squeezing pressure, 
the average interfacial separation depends 
logarithmically\cite{P4,P3,YP},
and the (projected) contact area linearly on the squeezing pressure\cite{P1}.
Here we study how the (average) interfacial separation depends on time when two elastic
solids with rough surfaces are squeezed together in a fluid. 

The influence of surface roughness on fluid flow at the interface between solids in stationary or sliding contact is
a topic of great importance both in nature and technology. Technological applications include leakage of seals,
mixed lubrication, and removal of water from the tire-road footprint. In nature fluid removal (squeeze-out) 
is important for adhesion and grip between the tree frog or gecko adhesive toe pads and the countersurface during rain,
and for cell adhesion. 

Almost all surfaces in nature and most surfaces of interest in tribology have roughness on many different length
scales, sometimes extending from atomic distances ($\sim 1 \ {\rm nm}$) to the macroscopic size of the system which could be of order
$\sim 1 \ {\rm cm}$. Often the roughness is fractal-like so that when a small region is magnified (in general with different
magnification in the parallel and orthogonal directions) it ``looks the same'' as the unmagnified surface. 

Most objects produced in engineering have some particular macroscopic shape characterized by a radius of curvature 
(which may vary over the surface of the solid) e.g., the radius $R$ of a cylinder in an engine. 
In this case the surface may appear perfectly smooth
to the naked eye but at short enough length scale, in general much smaller than $R$, the surface will exhibit strong irregularities
(surface roughness). The surface roughness power spectrum $C({\bf q})$ of such a surface exhibits a roll-off wavelength
$\lambda_0 << R$ (related to the roll-off wavevector $q_0=2 \pi /\lambda_0$) and therefore it appears smooth (except for the macroscopic curvature $R$)
on length scales much longer than $\lambda_0$. In this case, when studying the fluid flow between two macroscopic solids, one may
replace the microscopic equations of fluid dynamics with effective equations 
describing the average fluid flow on length scales much larger than $\lambda_0$, and which can be used to study, e.g., the lubrication of the
cylinder in an engine. This approach of eliminating or integrating out short length scale degrees of freedom to obtain effective
equations of motion, which describe the long distance (or slow) behavior is a very general and powerful concept often used in physics.
 
In the context of fluid flow at the interface between closely spaced solids with surface roughness, Patir and Cheng\cite{PC1,PC2} 
have showed how the Navier-Stokes equations of fluid dynamics
can be reduced to effective equations of motion involving locally averaged fluid pressures and flow velocities.
In the effective equation occurs so called flow factors, which are functions of the locally averaged interfacial
surface separation. They showed how the flow factors can be determined by solving numerically the fluid flow in small rectangular
units with linear size of order (or larger than) 
the roll-off wavelength $\lambda_0$ introduced above. In Ref. \cite{Pekl} one of us has developed an analytical theory for the
pressure flow factors based on the Persson contact mechanics model and the Bruggeman effective medium theory to take into account the
topography disorder resulting from the random roughness. We will use this theory in the calculations presented below. 

This paper is organized as follows: In Sec. 2 we briefly review the basic equations of fluid dynamics and describe
some simplifications which are valid for the present case. In Sec. 3 we describe the experimental method we have used to
study the interfacial separation and in Sec. 4 we compare the experimental results to the theory prediction. The summary and 
conclusions are presented in Sec. 5.

\begin{figure}
\includegraphics[width=0.45\textwidth,angle=0.0]{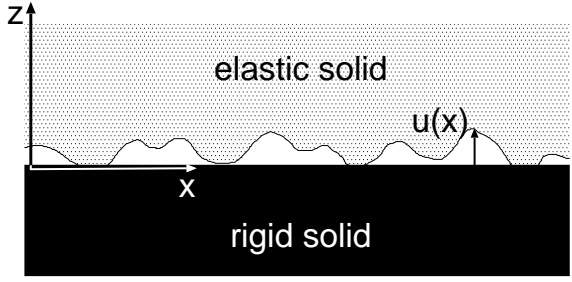}
\caption{\label{geometry}
An elastic solid with a rough surface in contact with a rigid solid with a flat surface.
}
\end{figure}

\vskip 0.3cm
{\bf 2. Theory}

\vskip 0.3cm
{\bf 2.1 Fluid flow between solids with random surface roughness}

Consider two elastic solids with randomly rough surfaces. Even if the solids are squeezed in contact,
because of the surface roughness there will in general be non-contact regions at the interface and, if the
squeezing force is not too large, there will exist non-contact channels from one side to the other side of the nominal
contact region. We consider now fluid flow at the interface between the solids. We assume that the fluid is Newtonian and
that the fluid velocity field ${\bf v}({\bf x},t)$ satisfies the Navier-Stokes equation:
$${\partial {\bf v} \over \partial t} + {\bf v}\cdot \nabla {\bf v} = -{1\over \rho} \nabla p + \nu \nabla^2 {\bf v}$$  
where $\nu = \eta /\rho$ is the kinetic viscosity and $\rho$ the mass density. For simplicity we will also assume
an incompressible fluid so that
$$\nabla \cdot {\bf v} = 0$$
We assume that the non-linear term ${\bf v}\cdot \nabla {\bf v}$ can be neglected (which corresponds to
small inertia and small Reynolds number), which is usually the case in fluid flow between narrowly spaced solid walls. 
For simplicity we assume the lower solid to be rigid with
a flat surface, while the upper solid is elastic with a rough surface. 
Introduce a coordinate system $xyz$ with the
$xy$-plane in the surface of the lower solid and the $z$-axis pointing towards the upper solid, see Fig. \ref{geometry}. 
The upper solid moves with
the velocity ${\bf v}_0$ parallel to the lower solid. 
Let $u(x,y,t)$ be the separation between the solid walls and assume that the slope $|\nabla u| << 1$. 
We also assume that $u/L << 1$, where $L$ is the linear size of the nominal contact region. In this case one expect that the fluid velocity varies slowly
with the coordinates $x$ and $y$ as compared to the variation in the orthogonal direction $z$.
Assuming a slow time dependence, the Navier Stokes equations reduces to
$$\eta {\partial^2 {\bf v} \over \partial z^2} \approx \nabla p$$
Here and in what follows ${\bf v} = (v_x,v_y)$, ${\bf x}=(x,y)$ and $\nabla = (\partial_x,\partial_y)$ are two-dimensional 
vectors. Note that $v_z \approx 0$ and that $p({\bf x})$ is independent of $z$ to a good approximation.
The solution to the equations above can be written as
$${\bf v} \approx {1\over 2 \eta} z(z-u({\bf x}))\nabla p + {z\over u({\bf x})}{\bf v}_0\eqno(1)$$
so that ${\bf v}=0$ on the solid wall $z=0$ and ${\bf v}={\bf v}_0$ for $z=u({\bf x})$.
Integrating over $z$ (from $z=0$ to $z=u({\bf x})$)
gives the fluid flow vector
$${\bf J} = - {u^3({\bf x})\over 12 \eta}\nabla p +  {1\over 2} u({\bf x}) {\bf v}_0\eqno(2)$$
Mass conservation demand that
$${\partial u({\bf x},t) \over \partial t} + \nabla \cdot {\bf J} = 0\eqno(3)$$ 
where the interfacial separation $u({\bf x},t)$ is the volume of fluid per unit area. In this last equation we have allowed
for a slow time dependence of $u({\bf x},t)$ as would be the case, e.g., during fluid squeeze-out from the 
interfacial region between two solids. 

\vskip 0.3cm
{\bf 2.2 Viscosity of confined fluids}

It is well known that the viscosity of fluids at high pressures may be many orders of magnitude larger than at low pressures.
Using the theory of activated processes, and assuming that a local molecular rearrangement in a fluid results in a local volume expansion,
one expect an exponential dependence on the hydrostatic pressure $\eta = \eta_0 {\rm exp}(p/p_0)$, where typically
(for hydrocarbons or polymer fluids) $p_0 \approx 10^8 \ {\rm Pa}$. Here we are interested in
(wetting) fluids confined between the surfaces of elastically soft solids, e.g., rubber or gelatin. In this case the pressure at the interface
is usually at most of order the Young's modulus, which (for rubber) is less than $10^7 \ {\rm Pa}$. 
Thus, in most cases involving elastically soft materials, the viscosity 
can be considered as independent of the local pressure. 
In the applications below
the nominal pressure is only of order $\sim 10^4 \ {\rm Pa}$ and the pressure in the area of real contact of order $\sim 10^6 \ {\rm Pa}$,
so that the dependence of the (bulk) viscosity on the pressure can be neglected. 
Nevertheless, it has been observed experimentally\cite{eta1,eta2},
and also found in Molecular Dynamics (MD) simulations\cite{theory1a,theory1b}, that the effective viscosity $\eta$
(defined by $\sigma = \eta v/u$, where $\sigma $ is the shear stress, $u$ the separation between the surfaces 
and $v$ the relative velocity) of very thin (nanometer thickness) fluid films
confined between solid walls at low pressure may be strongly enhanced, and to exhibit non-Newtonian properties. 
In addition, for nanometer wall-wall separations, a finite normal stress is necessary for squeeze-out, i.e., the ``fluid'' now
behaves as a soft solid and the squeeze-out occurs in a quantized way by removing one monolayer after another
with increasing normal stress\cite{theory2}. 
In the application below we only study the (average) separation between the walls
with micrometer resolution, and in this case the strong increase in
the viscosity for very short wall separations becomes irrelevant. 

\vskip 0.3cm
{\bf 2.3 Roughness on many length scales: effective equations of fluid flow}

\begin{figure}
\includegraphics[width=0.45\textwidth,angle=0.0]{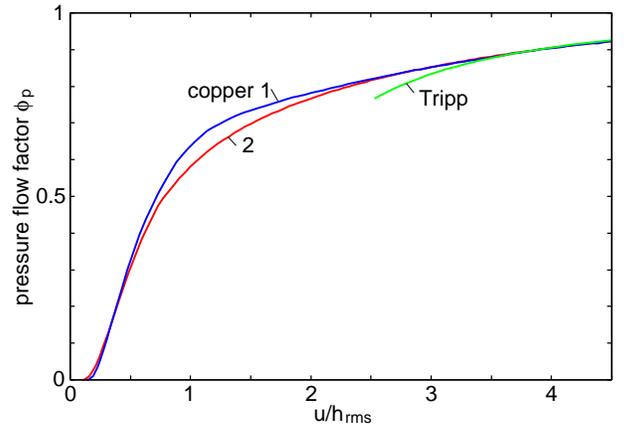}
\caption{\label{flowfactor}
The fluid pressure flow factor as a function of the
average interfacial separation $u$ divided by the root-mean-square roughness amplitude $h_{\rm rms}$.
For copper surfaces {\bf 1} and {\bf 2}.
The green curve shows the large $u$-behavior predicted by Tripp\cite{Tripp}.
}
\end{figure}

Equations (2) and (3) describe the fluid flow at the interface between contacting solids with rough surfaces.
The surface roughness can be eliminated or integrated out using the Renormalization Group (RG) procedure.
In this procedure one eliminate or integrate out the surface roughness components in steps and obtain a set of RG flow equations describing how the
effective fluid equation evolves as more and more of the surface roughness components are eliminated.
One can show that after eliminating all the surface roughness components,  
the fluid current [given by (2)] takes the form
$${\bf J} = A(u) \nabla p + B(u) {\bf v}\eqno(4)$$
where $A$ and $B$ are $2\times 2$ matrices, and where $u({\bf x},t)$ and $p({\bf x},t)$ now are {\it locally averaged}
quantities. In general, $A$ and $B$ depend also on $\nabla p$ (see Ref. \cite{mic}), but for the low pressures (and pressure gradients)
prevailing in the application presented below, we can neglect this effect. 
If the sliding velocity ${\bf v}=0$ and if the surface roughness has isotropic statistical
properties, then $A$ is proportional to the unit matrix and is usually written as $A= - u^3 \phi_{\rm p} ( u)/ (12 \eta)$.
In this case from (3) and (4) we get
$${\partial u \over \partial t} - \nabla \cdot \left ({u^3 \phi_{\rm p} ( u) \over 12 \eta} \nabla p\right ) = 0\eqno(5)$$ 
If $u({\bf x},t)$ is independent of ${\bf x}$ then (5) takes the form
$${d u \over d t} -{ u^3 \phi_{\rm p} (u)\over 12 \eta} \nabla^2 p  = 0\eqno(6)$$ 
In Fig. \ref{flowfactor} we show $\phi_{\rm p} (u)$ calculated using the Persson contact mechanics
and the Bruggeman effective medium theory\cite{Pekl}. The figure shows the 
dependence of $\phi_{\rm p} (u)$ on the separation $u$ for the two (copper) 
surfaces used in the study below.
The green curve shows the large $u$-behavior predicted by Tripp\cite{Tripp}:
$$\phi_{\rm p} \approx 1-{3\over 2} {\langle h^2 \rangle \over u^2 }$$
where $\langle h^2 \rangle = h^2_{\rm rms}$. See also Ref. \cite{Pekl} for the calculation of higher order corrections.
The small contact pressure involved in the experiments reported on below 
results in relative large (average) separation between the surfaces, $u > 1.4 h_{\rm rms}$ on both surfaces
(as calculated using the theory developed in Ref. \cite{P4,YP}).

\vskip 0.3cm
{\bf 2.4 Fluid squeeze-out}

Let us squeeze a cylindrical rubber block (height $d$ and radius $R$) against a substrate in a fluid.
Assume that we can neglect the macroscopic deformations of the rubber block in response to the (macroscopically)
non-uniform fluid pressure. In this case $u({\bf x},t)$ will only depend on time $t$ and (6) will hold. 
Equation (6) imply that the fluid pressure
$$p = 2 p_{\rm fluid} \left (1-{r^2\over R^2}\right )\eqno(7)$$
where $r=|{\bf x}|$ denote the distance from the cylinder axis, and
where we have assumed that the external pressure vanish. $p_{\rm fluid}$ denotes the {\it average}
fluid pressure in the nominal contact region. Substituting (7) in (6) gives
$${d u\over dt} \approx -{2 u^3 \phi_{\rm p}( u) p_{\rm fluid} (t) \over 3 \eta R^2}\eqno(8)$$

If $p_0$ is the applied pressure acting on the top surface of the cylinder block,
we have
$$p_{\rm fluid} (t) = p_0 - p_{\rm cont}(t), \eqno(9)$$
where $p_{\rm cont}$ is the asperity contact pressure. 
We first assume that the pressure $p_0$ is so small that for all times 
$u >> h_{\rm rms}$ and in this case 
$\phi_{\rm p} (u) \approx 1$. For $u >> h_{\rm rms}$ we also have\cite{P4}
$$p_{\rm cont} \approx \beta  
E^* {\rm exp}\left ( - {u \over u_0}\right ).\eqno(10)$$
where $E^* = E/(1-\nu^2)$ (where $E$ is the Young modulus and $\nu$ the Poisson ratio), and 
where $u_0 = h_{\rm rms}/\alpha$. The parameters $\alpha$ and $\beta$ depends on the fractal properties 
of the rough surface\cite{P4}. 
Using (10) and (9) we get from (8):
$${d p_{\rm cont} \over dt} \approx {2 u^3 (p_{\rm cont}(t)) \over 3 \eta R^2 u_0}p_{\rm cont}\left (p_0-p_{\rm cont}\right ),\eqno(11)$$
For long times $p_{\rm cont} \approx p_0$ and  
we can approximate (11) with
$${d p_{\rm cont} \over dt} \approx {2  u^3 (p_0) \over 3 \eta R^2 u_0} p_0 (p_0-p_{\rm cont}).$$
Integrating this equation gives
$$p_{\rm cont} (t) \approx p_0 - \left [p_0-p_{\rm cont}(0)\right ] {\rm exp} 
\left (-\left ({ u (p_0)\over h_{\rm rms}}\right )^3 {t\over \tau} \right )$$
where
$$\tau = {3\eta R^2 u_0 \over 2 h^3_{\rm rms} p_0} = 
{3 \eta R^2  \over 2 \alpha h^2_{\rm rms} p_0}.\eqno(12)$$
Using (12) this gives
$$u \approx  u_\infty + \left ( 1-{p_{\rm cont}(0)\over p_0} \right ) 
u_0 {\rm exp} \left ( -\left ({ u (p_0)\over h_{\rm rms}}\right )^3 {t\over \tau} \right )$$
where $u_\infty = u_0 {\rm log}(\beta E^*/p_0)$.
Thus, $u(t)$ will approach the equilibrium separation $u_\infty$ in an exponential way, and
we can define the squeeze-out time as the time to reach, say, $1.01 u_\infty$, which typically will be a few times
$\tau' =  [h_{\rm rms}/u(p_0)]^{3} \tau $. In the experiments performed below 
$\eta = 100 \ {\rm Pas}$, $R \approx 1 \ {\rm cm}$ and $p_0 \approx 10^4 \ {\rm Pa}$.
Using that $u_0 \approx h_{\rm rms}$ and that for the rough surfaces used below $h_{\rm rms} \approx 50 \ {\rm \mu m}$,
and $u(p_0) \approx 1.4 h_{\rm rms}$ we get $\tau' \approx 1000 \ {\rm s}$. Thus we expect squeeze-out to occur in 
about 1 hour, in good agreement with the experimental data (see below). 
For flat surfaces, within continuum mechanics, the film thickness approaches zero as $t\rightarrow \infty$ as
$ u \sim t^{-1/2}$. Thus in this case there is no natural or characteristic time-scale, and  
it is not possible to define a meaningful fluid squeeze-out time.


At high enough squeezing pressures and after long enough time, the interfacial separation will
be smaller than $h_{\rm rms}$, so that
the asymptotic relation (10) will no longer hold. In this case the relation
$p_{\rm cont} ( u )$ can be calculated using the equations given in Ref. \cite{Sealing}. 
Substituting (9) in (8) and measuring pressure in units of
$p_0$, separation in units of $h_{\rm rms}$ and time in units of $\tau$ one obtain 
$${d u\over dt} \approx -\alpha^{-1}  \phi_{\rm p}(u) u^3 (1-p_{\rm cont}),\eqno(13)$$
where $\alpha = h_{\rm rms}/u_0$.
This equation together with the relation $p_{\rm cont} ( u )$ constitute
two equations for two unknown ($ u$ and $p_{\rm cont}$) which can be easily solved
by numerical integration. 

\begin{figure}[tbp]
\includegraphics[width=0.48\textwidth,angle=0]{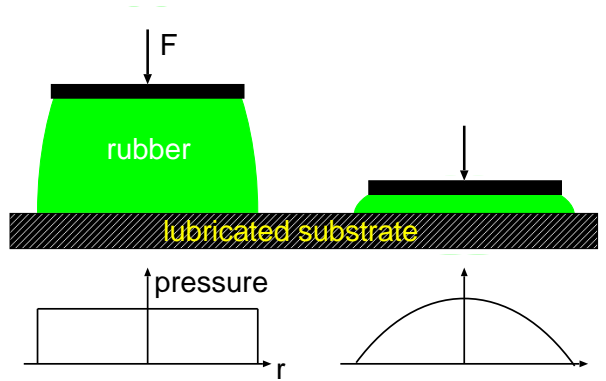}
\caption{
A cylindrical rubber block (height $d$ and radius $R$) squeezed against a lubricated substrate
(no friction). If $d>R$ the pressure distribution at the interface will be nearly uniform (left) while
if $d<<R$ (right) the pressure distribution will be nearly parabolic. We have assumed that
the upper surfaces of the rubber cylinders are glued (no slip) to a flat rigid disk. 
} 
\label{twocylinderlub}
\end{figure}

\begin{figure}[tbp]
\includegraphics[width=0.35\textwidth,angle=0]{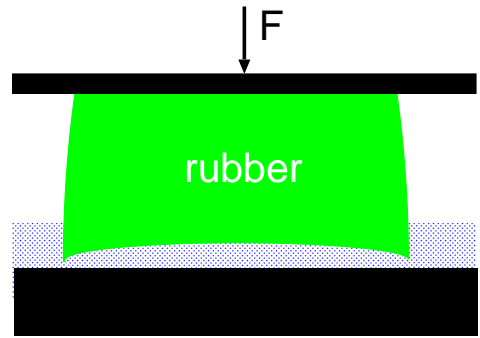}
\caption{
The non-uniform hydrodynamic pressure is highest at the center of the contact region and will deform the rubber
block as indicated in the figure.}
\label{picdeformed}
\end{figure}

\vskip 0.3cm
{\bf 2.5 Rubber block under vertical loading}

Consider a cylindrical rubber block (radius $R$, height $d$) squeezed between two flat surfaces. If both surfaces are lubricated
(no friction) the stress at the interfaces will be constant $p=p_0 = F_{\rm N}/\pi R^2$ and the change in
the thickness $\Delta d$ (assuming linear elasticity) will be determined by $p_0 = E \Delta d / d$.
However, if the rubber adhere (or is glued) to the upper surface with no-slip the situation may be very different\cite{Gent,Thesis}.
If $d > R$ the stress at the (lower) interface will again be nearly uniform and $\Delta d$ will be determined by
$p_0 = E_{\rm eff} \Delta d / d$ where $E_{\rm eff} >E$ but nearly identical to $E$. In the opposite limit of
a very thin rubber disk, $d << R$, the pressure distribution at the bottom surface will be nearly
parabolic $p(r) \approx 2 p_0 [1-(r/R)^2]$ (see Fig. \ref{twocylinderlub}) and the effective elasticity $E_{\rm eff} >> E$.
Experiments have shown that when a rubber disk is squeezed against a rough surface, even if the rubber disk is very thin,
the (locally averaged) pressure distribution at the bottom surface of the rubber disk will be nearly constant\cite{newexp}. 
This is because the rubber is pressed into the ridges on the rough surface
under vertical loading, and the hydrostatic pressure becomes smaller. We also note that while $E_{\rm eff} > E$ determines the 
change in the thickness of the rubber block, the local elastic asperity-induced deformations 
at the (lower) interface will be determined (to a good approximation) by the Young's modulus $E$. 

In Sec. 2.4 we have shown that a flat cylinder surface squeezed against a flat substrate in a (Newtonian) fluid
gives rise to a parabolic fluid pressure distribution. This implies that for a very thin ($d << R$) elastic disk,
glued to a flat rigid surface, and squeezed against another flat surface in a fluid,
we expect the bottom surface of the elastic disk to remain nearly flat and the assumption made in 
Sec. 2.4 will hold to good accuracy. However, if the rubber block is thick enough ($d >R$) the bottom surface of the block
will bend inwards as indicated in Fig. \ref{picdeformed}, which will slow down the fluid squeeze-out.

\begin{figure}[tbp]
\includegraphics[width=0.45\textwidth,angle=0]{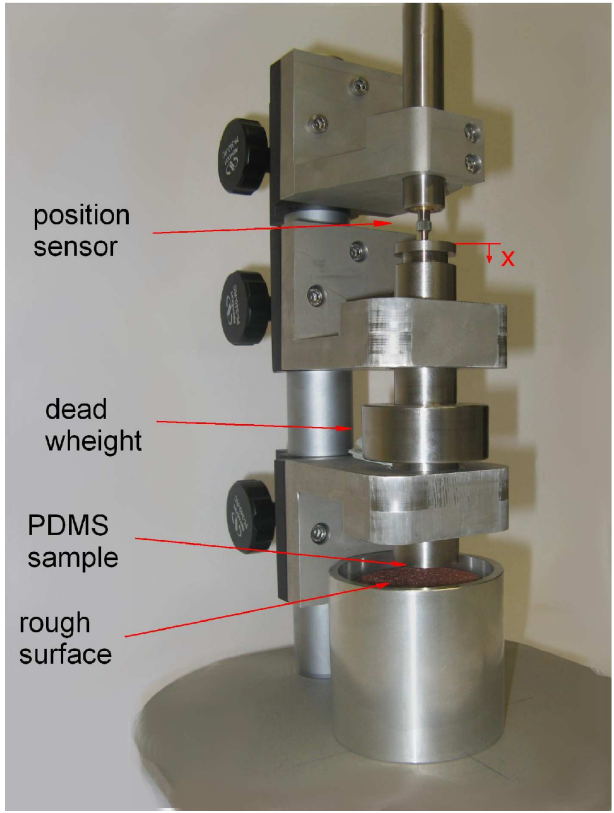}
\caption{
Experimental set-up for studies of fluid squeeze-out between surfaces of elastic solids.}
\label{setup}
\end{figure}

\vskip 0.3cm
{\bf 3. Experimental}

We have studied the squeeze-out of fluids between solids with rough surfaces as shown in Fig. \ref{setup}. 
In the experimental set-up a cylindrical silicon rubber block is squeezed against a rough counter-surface in the presents of a fluid. 
The rubber block is attached to a dead weight, resulting in the driving force $F_{\rm N} = 13.8 \ {\rm N}$. 
This force is kept constant for all experiments. We have measured the downwards movement of the dead weight as a function of 
time using a digital gauge with a relative position resolution of $0.5 \ {\rm \mu m}$. In order to slow down the 
whole process, we use a very high viscosity silicon oil 
(Dow Corning 200 Fluid, viscosity $100 \ {\rm Pa s}$) and a relative low nominal squeezing pressure (about $10^4$ Pa).
In the different configurations we either squeeze an elastic silicon rubber block, or a rigid glass block, against smooth (glass) 
or rough (copper) surfaces in order to test different aspects of the squeeze-out. The rubber blocks have the radius $R=1.5 \ {\rm cm}$ and 
height $d=1 \ {\rm cm}$, $0.5 \ {\rm cm} $ and $0.3 \ {\rm cm}$. We use a silicone elastomer (PDMS) prepared using a two-component kit (Sylgard 184) 
purchased from Dow Corning (Midland, MI). This kit consists of a base (vinyl-terminated polydimethylsiloxane) and a curing agent 
(methylhydrosiloxane.dimethylsiloxane copolymer) with a suitable catalyst. 
From these two components we prepared a mixture 10:1 (base/cross linker) in weight. 
The mixture was degassed to remove the trapped air induced by stirring from the mixing 
process and then poured into cylindrical casts. The bottom of these casts 
was made from glass to obtain smooth surfaces (negligible roughness). The samples were cured in an oven at $80^\circ{\rm C}$  for over 12 hours.
The rough copper surfaces where prepared by pressing sandpaper surfaces against flat and plastically soft copper surfaces using a hydraulic press. 
Using sandpaper with different grit size, and repeating the procedure many times, resulted in (nearly) randomly rough 
surfaces suitable for our experiment.

The silicon block was placed in the high viscosity fluid with some distance to the rough surface. 
In order to avoid kinetic (inertia) effects the initial separation was selected to be very small. The nominal force was applied by dropping the dead weight 
with the rubber block attached to it. The displacement of the dead weight from its starting position 
was measured as a function of time.

\begin{figure}[tbp]
\includegraphics[width=0.45\textwidth,angle=0]{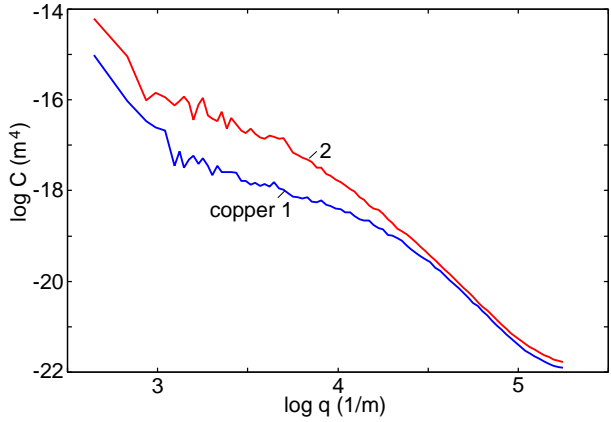}
\caption{
The logarithm of the surface roughness power spectrum as a function of the  logarithm of the wavevector
for two copper surfaces, ${\bf 1}$ and ${\bf 2}$, with the root-mean-square 
roughness $42 \ {\rm \mu m}$ and $88 \ {\rm \mu m}$, respectively.} 
\label{Cq.copper1.2}
\end{figure}

\begin{figure}[tbp]
\includegraphics[width=0.45\textwidth,angle=0]{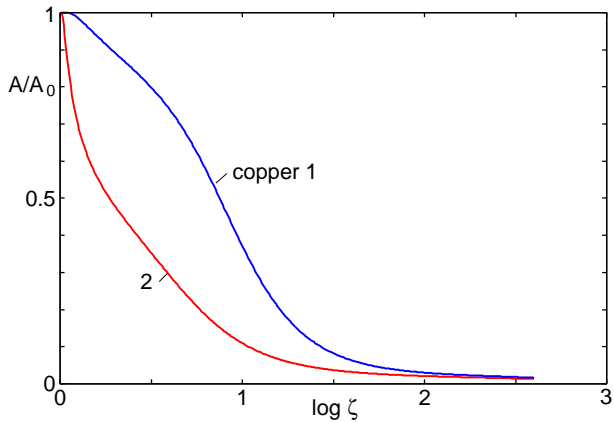}
\caption{
The calculated relative area of contact $A/A_0$ (where $A_0$ is the nominal contact area)
when a glass and a PDMS 
block is squeezed against two copper surfaces ${\bf 1}$ and ${\bf 2}$ with surface roughness produced as described in the text.
} 
\label{logzeta.Area.Cu1.Cu2}
\end{figure}

In Fig. \ref{Cq.copper1.2} we show the power spectrum of the two rough copper surfaces ${\bf 1}$ and ${\bf 2}$ used in our study. The area of real contact
(at the nominal squeezing pressure $\approx 2\times 10^4 \ {\rm Pa}$) as a function of the magnification $\zeta$ 
is shown in Fig. \ref{logzeta.Area.Cu1.Cu2}.
Note that the area of real contact (i.e., the contact area at the highest magnification $\zeta_1$ or wavevector $q_1=q_0\zeta_1$)  
is rather similar in both cases 
(equal to $A = 0.016 A_0$ and $0.013 A_0$ for surfaces ${\bf 1}$ and ${\bf 2}$, respectively) in spite of the rather large difference
in the rms roughness values ($h_{\rm rms}=42$ and $88 \ {\rm \mu m}$, respectively). 
This is due to the fact that the rms roughness is dominated by the longest wavelength roughness components,
while the area of real contact is strongly influenced by the short wavelength roughness components,
which are very similar on both surfaces (see Fig. \ref{Cq.copper1.2} for large wavevector).
The small contact pressure result in relative large (average) separation between the surfaces, $\bar u \approx 1.4 h_{\rm rms}$ on both surfaces
(as calculated using the theory developed in Ref. \cite{P4,YP}).

\begin{figure}[tbp]
\includegraphics[width=0.45\textwidth,angle=0]{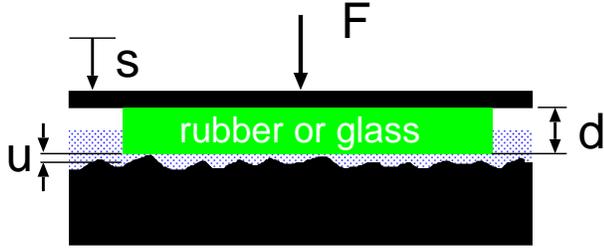}
\caption{
Squeeze-out experimental set-up. A cylindrical 
glass or rubber block is squeezed against a substrate with a smooth or rough surface
in a fluid. The cylindrical body has the height $d= 0.5 \ {\rm cm}$ thick and the diameter $D=2R = 3 \ {\rm cm}$.
The normal load $F_{\rm N} = 13.8 \ {\rm N}$ and the fluid viscosity $\eta = 100 \ {\rm Pa s}$. 
The vertical displacement $u$ of the upper surface is registered as a function of time.}
\label{squeezeoutpic}
\end{figure}

\begin{figure}[tbp]
\includegraphics[width=0.45\textwidth,angle=0]{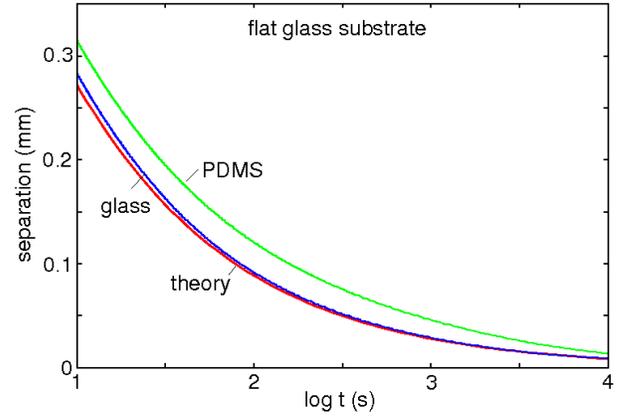}
\caption{
The surface separation as a function of the logarithm of time when a glass and a PDMS cylindrical
block is squeezed against a flat glass substrate in a silicon oil. Also shown is the theoretical prediction
(lower curve). The cylindrical body has the height $d= 0.5 \ {\rm cm}$ thick and has the diameter $D=2R = 3 \ {\rm cm}$.
The normal load $F_{\rm N} = 13.8 \ {\rm N}$ and the fluid viscosity $\eta = 100 \ {\rm Pa s}$.}
\label{flat.glass.extended.1logtime.2h}
\end{figure}



\begin{figure}[tbp]
\includegraphics[width=0.45\textwidth,angle=0]{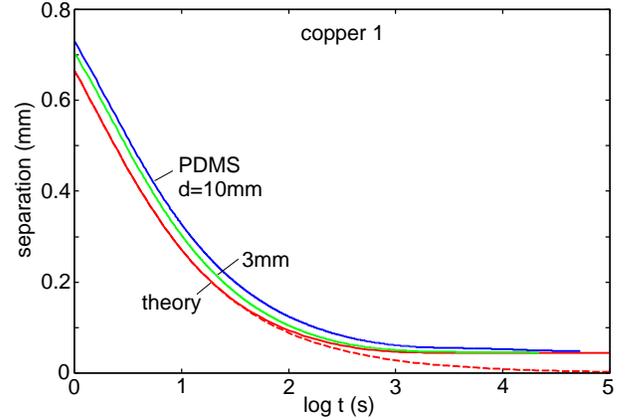}
\caption{
The surface separation as a function of the logarithm of time when PDMS cylindrical
blocks with thickness $10 \ {\rm mm}$ and $3 \ {\rm mm}$ are squeezed against a rough copper 
surface ${\bf 1}$ (root-mean-square roughness $h_{\rm rms} = 42 \ {\rm \mu m}$) 
in a silicon oil. Also shown is the theoretical prediction for a flat substrate (dashed curve) and for the copper surface ${\bf 1}$
(lower solid curve).} 
\label{copper1.1time.2h}
\end{figure}

\begin{figure}[tbp]
\includegraphics[width=0.45\textwidth,angle=0]{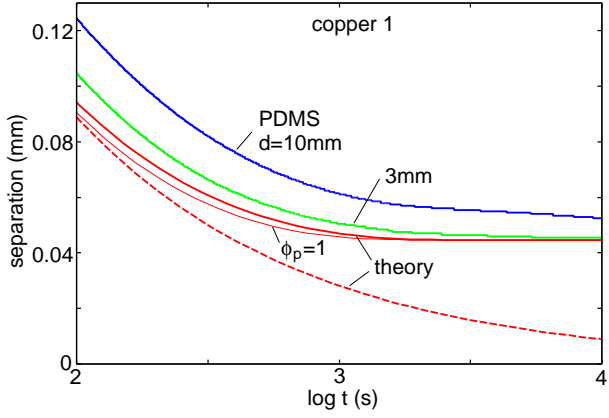}
\caption{
The same as in Fig. \ref{copper1.1time.2h} but for a more narrow time interval.
The thin solid line is the calculated squeeze-out when the pressure flow factor $\phi_p(u)=1$}
\label{copper1.1time.2h.mag}
\end{figure}

\begin{figure}[tbp]
\includegraphics[width=0.45\textwidth,angle=0]{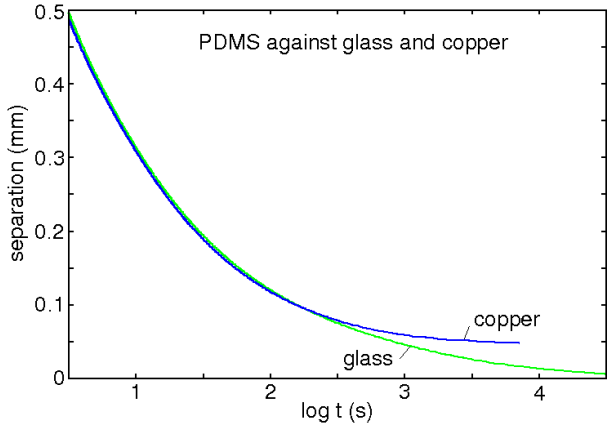}
\caption{
The surface separation as a function of the logarithm of time when a $d=0.5 \ {\rm cm}$ thick PDMS cylindrical
block is squeezed against a flat glass substrate (lower curve), and 
against the (rough) copper surface ${\bf 1}$ (root-mean-square roughness $h_{\rm rms} = 42 \ {\rm \mu m}$) 
in a silicon oil.} 
\label{PDMS.glass.copper}
\end{figure}

\begin{figure}[tbp]
\includegraphics[width=0.45\textwidth,angle=0]{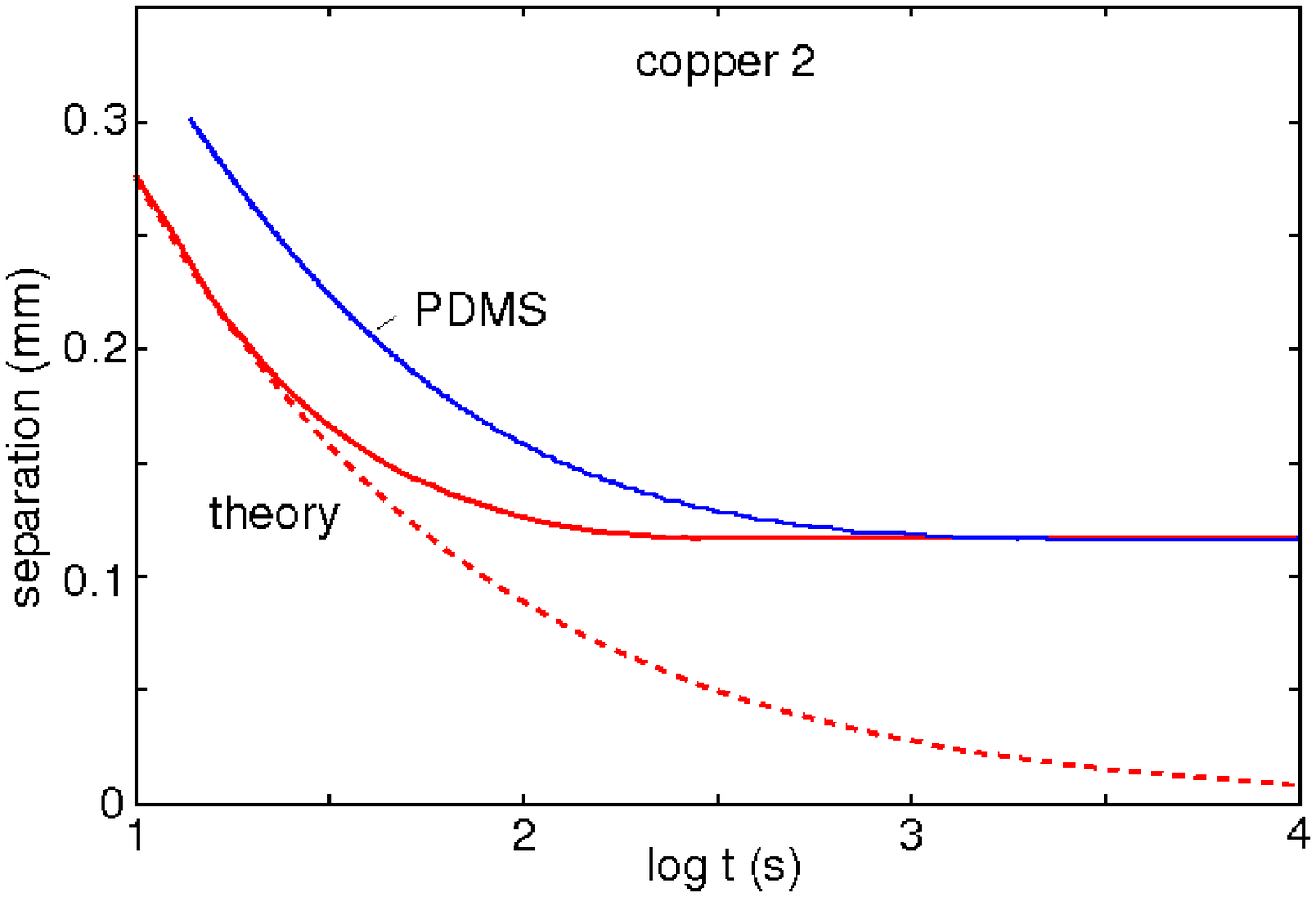}
\caption{
The surface separation as a function of the logarithm of time when a $d=0.5 \ {\rm cm}$ thick PDMS cylindrical
block is squeezed against the rough copper surface ${\bf 2}$ (root-mean-square roughness $h_{\rm rms} = 88 \ {\rm \mu m}$) 
in a silicon oil. Also shown is the theoretical prediction
(lower curve).} 
\label{copper2.1logtime.2h}
\end{figure}

\vskip 0.3cm
{\bf 4. Comparison of theory with experiment}

In Fig. \ref{flat.glass.extended.1logtime.2h} we show
the surface separation as a function of the logarithm of time when a glass and a PDMS cylindrical
block are squeezed against a flat glass substrate in a silicon oil. Also shown is the theoretical prediction
(lower curve). The cylinder is $d= 0.5 \ {\rm cm}$ thick and has the diameter $D=2R = 3 \ {\rm cm}$.
As expected, the theory result agree almost perfectly with the experimental results 
for the glass cylinder  (no fitting parameters), but for the rubber
block the (average) separation is larger and the squeeze-out slower. We attribute this to 
temporarily trapped fluid resulting from the upward bending (before contact with the substrate) of the bottom surface of the rubber
block as in Fig. \ref{picdeformed}.  
We define the ``trapped'' fluid volume $\Delta V$ as the fluid volume between the bottom surface of the
block and a flat (mathematical) surface in contact with the block at the edge $r=R$ of the bottom surface of the block. 
Using the theory of elasticity $\Delta V = \pi R^2 \delta$ with $\delta = C R \bar p/E_{\rm eff}$, where
$C$ is a constant of order unity. 
For the rubber block with thickness $d = 0.5 \ {\rm cm}$ we 
get $E_{\rm eff} \approx E[1+0.5\times (R/2d)^2]\approx 4 \ {\rm MPa}$ and $\delta \approx 40 \ {\rm \mu m}$,
resulting in an increase in the average interfacial separation by $\sim 40 \ {\rm \mu m}$, which is consistent with what we observe. 

Fig. \ref{copper1.1time.2h} and 
\ref{copper1.1time.2h.mag} 
show the surface separation as a function of the logarithm of time when PDMS cylindrical
blocks with thickness $1 \ {\rm cm}$ and $0.3 \ {\rm cm}$ are squeezed against the rough copper 
surface ${\bf 1}$ (root-mean-square roughness $h_{\rm rms} = 42 \ {\rm \mu m}$) 
in a silicon oil. 
Also shown is the theoretical prediction for a flat substrate (dashed curve) and for the copper surface
(lower solid curve), assuming that the bottom surface of the rubber disk is macroscopically
flat. Note that the agreement between the theory and experiment is much better for the thinner rubber
disk. This is indeed expected since the fluid-pressure induced curvature of the bottom surface of the rubber is smaller for the thin rubber
disk (see Sec. 2.5). But even for the thin rubber disk some fluid-pressure induced bending of the bottom surface 
of the rubber disk is expected, and we believe this is the main origin for the slightly slower squeeze-out observed in the experiment as
compared to the theory prediction.
The thin solid line Fig. \ref{copper1.1time.2h.mag} shows the calculated squeeze-out when the pressure flow factor $\phi_p(u)=1$.
In the present case the pressure flow factor is close to unity and this explains the relative small difference between using
$\phi_p=1$ (thin red line) and using the calculated $\phi_p(u)$ (from Fig. \ref{flowfactor}) (thick red line).

In Fig. \ref{PDMS.glass.copper}
we show the surface separation as a function of the logarithm of time when the $d= 0.5 \ {\rm cm}$ thick PDMS disk 
is squeezed against a flat glass substrate (lower curve), and 
against the (rough) copper surface {\bf 1} (root-mean-square roughness $h_{\rm rms} = 42 \ {\rm \mu m}$) 
in a silicon oil.  Note that before contact with the substrate, the fluid-pressure induced bending of the bottom
surface of the block is the same in both cases (giving overlapping curves for $t < 300 \ {\rm s}$).

In Fig. \ref{copper2.1logtime.2h}
we show the surface separation as a function of the logarithm of time when the $d=0.5 \ {\rm cm}$ thick PDMS cylindrical
block is squeezed against the (rough) copper surface ${\bf 2}$ (root-mean-square roughness $h_{\rm rms} = 88 \ {\rm \mu m}$) 
in a silicon oil. Also shown is the theoretical prediction
(lower curve). Again, the bending of the bottom surface of the rubber block results in a slower squeeze-out than predicted
theoretically assuming a (macroscopically) flat bottom surface of the rubber block.

\vskip 0.3cm
{\bf 5. Summary and conclusion}

In this paper we have studied the fluid squeeze-out from the interface between an 
elastic block with a flat surface and a randomly rough surface of a rigid solid. 
We have calculated the (average) interfacial separation 
as a function of time by considering the fluid flow using a contact mechanics theory
in combination with thin-film hydrodynamics with flow factors (which are functions 
of the (local) interfacial separation) obtained using a recently developed theory. 
We have explained the importance of the large length-scale elastic deformations on the squeeze out.

The theoretical results have been compared to experimental results. The experiment was performed
by squeezing cylindrical rubber blocks with different height $d$ against rough cooper surfaces in the 
presents of a high viscosity fluid (silicone oil). Changing the height $d$ of the rubber block, 
and also performing additional experiments with flat against flat surfaces, with combinations of 
rigid-rigid and elastic-rigid, we could show the importance of both the large length-scale and asperity induced
elastic deformation on the squeeze-out. In particular, large length-scale deformations of the bottom surface of the rubber
block resulted in (temporary) trapped fluid between the elastic solid and the rigid countersurface, which drastically
slowed-down the squeeze-out. This effect is smallest for the thinnest rubber block, for which case we find  
good agreement between the theory (where we have neglected the large length-scale deformations of the rubber block)
and the experiments. Another mechanism which drastically slows down the squeeze-out occurs at much higher nominal
pressure (or load) than used in the present experiment. 
This is due to sealed-off fluid in the nominal contact region during contaqct formation.
This effect occurs when the area of real contact approaches $\approx 0.4 A_0$, where
the area of real contact percolate resulting in sealed-off regions of fluid, which may disappear only extremely
slowly, e.g., by diffusion into the rubber. This effect was discussed in Ref. \cite{squeezeout} and seams to be of importance in
many applications involving high contact pressures, e.g., it may result in a static (or start-up) friction force
which slowly increases with time even after very long time (say one year). 

The squeeze-out of fluids from the interfacial region between elastic solids with rough surfaces is very important in many 
technical applications (e.g. a tires rolling on a wet road, wipers and dynamic seals),
and the results presented in this paper contribute to this important subject.

\vskip 0.3cm

{\bf Acknowledgments}

This work, as part of the European Science Foundation EUROCORES Program FANAS, was supported from funds 
by the DFG and the EC Sixth Framework Program, under contract N ERAS-CT-2003-980409.


\begin{thebibliography}{999}

\bibitem{Sealing}
B.N.J. Persson and C. Yang, 
J. Phys.: Condens. Matt. {\bf 20}, 315011 (2008)

\bibitem{Capillary.adhesion}
B.N.J. Persson, 
J. Phys.: Condens. Matt. {\bf 20}, 315007 (2008)

\bibitem{Bowden}
F.P. Bowden and D. Tabor, {\it Friction and Lubrication of Solids}
(Wiley, New York, 1956).

\bibitem{Johnson}
K.L. Johnson, {\it Contact Mechanics},
(Cambridge University Press, Cambridge, 1966).

\bibitem{BookP}
B.N.J. Persson,
{\it Sliding Friction: Physical Principles and Applications}, 
2nd edn. (Springer, Heidelberg, 2000).

\bibitem{Isra}
J.N. Israelachvili,
{\it Intermolecular and Surface Forces} (Academic, London (1995)).

\bibitem{P33}
See, e.g., B.N.J. Persson, O. Albohr, U. Tartaglino, A.I. Volokitin and E. Tosatti, 
J. Phys. Condens. Matter {\bf 17}, R1 (2005). 

\bibitem{P3}
B.N.J. Persson, Surface Science Reports {\bf 61}, 201 (2006).

\bibitem{P4}
B.N.J. Persson, Phys. Rev. Lett. {\bf 99}, 125502 (2007) 

\bibitem{YP}
C. Yang and B.N.J. Persson, J. Phys. Condens. Matter {\bf 20}, 215214 (2008). 

\bibitem{P1}
B.N.J. Persson, J. Chem. Phys. {\bf 115}, 3840 (2001).

\bibitem{PC1}
N. Patir and H.S. Cheng, Journal of Tribology, Transactions of the ASME {\bf 100}, 12 (1978).

\bibitem{PC2}
N. Patir and H.S. Cheng, Journal of Tribology, Transactions of the ASME {\bf 101}, 220 (1979).

\bibitem{Pekl}
B.N.J. Persson, J. Phys.: Condens. Matter {\bf 22}, 265004 (2010).

\bibitem{eta1}
S. Yamada, Tribology Letters {\bf 13}, 167 (2002).

\bibitem{eta2}
L. Bureau,
Phys. Rev. Lett. {\bf 104}, 218302 (2010).

\bibitem{theory1a}
P.A. Thompson, G.S. Grest and M.O. Robbins, Phys. Rev. Lett. {\bf 68}, 3448 (1992).

\bibitem{theory1b}
I.M. Sivebaek, V.N. Samoilov and B.N.J. Persson, in preparation.

\bibitem{theory2}
B.N.J. Persson and F. Mugele, J. Phys.: Condens. Matter {\bf 16}, R295 (2004).

\bibitem{mic}
M. Scaraggi, G. Carbone, B.N.J. Persson and D. Dini, subm. to Soft Matter. 

\bibitem{Tripp}
J.H. Tripp, ASME J. Lubrication Technol. {\bf 105}, 485 (1983). 


\bibitem{Gent}
A.N. Gent, Rubber Chem. Technol., {\bf 67}, 549 (1994). 

\bibitem{Thesis}
J.B. Suh, {\it Stress analysis of rubber blocks under vertical and shear loading}, 
PhD thesis (2007).

http://etd.ohiolink.edu/view.cgi?akron1185822927

\bibitem{newexp}
E.A. Sakai, Tire Science and Technology {\bf 23}, 238 (1995).

\bibitem{squeezeout}
B. Lorenz and B.N.J. Persson, European Journal of Physics E{\bf 32}, 281 (2010).

\end{thebibliography}
\end{document}